\documentclass[iop,amsmath,amssymb,amsfonts,preprint,superscriptaddress,citeautoscript,byrevtex]{revtex4-1}
\usepackage{graphicx,color}
\usepackage{epstopdf}
\pdfoutput=1 
\usepackage{soul}
\usepackage{ulem}

\begin{document}
\title{{\it Ab-initio} study of electronic and magnetic properties of Mn$_2$RuZ/MgO (001) heterojunctions (Z= Al, Ga, Si, Ge)}
%\author{Tufan Roy$^{1}$\footnote{Electronic mail: tufanroyburdwan@gmail.com}, Masahito Tsujikawa$^{1,2}$,  Masafumi Shirai$^{1,2,3}$ }
%\affiliation {$^{1}$Research Institute of Electrical Communication, Tohoku University, Sendai 980-8579, Japan}
%\affiliation{$^{2}$ Center for Spintronics Research Network (CSRN), Tohoku University, Sendai 980-8579, Japan}
%\affiliation{$^{3}$ Center for Science and Innovation in Spintronics (CSIS), Core Research Cluster (CRC), Tohoku University, Sendai 980-8577, Japan}
\author{Tufan Roy}
\email{tufan@riec.tohoku.ac.jp}
\affiliation{Research Institute of Electrical Communication (RIEC), Tohoku University, Sendai 980-8577, Japan}
\author{Masahito Tsujikawa}
\affiliation{Research Institute of Electrical Communication (RIEC), Tohoku University, Sendai 980-8577, Japan}
\affiliation{Center for Spintronics Research Network (CSRN), Tohoku University, Sendai 980-8577, Japan}
\author{Masafumi Shirai}
\affiliation{Research Institute of Electrical Communication (RIEC), Tohoku University, Sendai 980-8577, Japan}
\affiliation{Center for Spintronics Research Network (CSRN), Tohoku University, Sendai 980-8577, Japan}
\affiliation{Center for Science and Innovation in Spintronics (CSIS), Core Research Cluster (CRC), Tohoku University, Sendai 980-8577, Japan}
%\date{\today}
\begin{abstract}
Using first-principles calculations, we studied Mn$_2$RuZ (Z=Al, Ga, Si, Ge) and their heterojunctions with MgO along (001) direction. All these alloys possess Hg$_2$CuTi-type inverse Heusler alloy structure and ferrimagnetic ground state. Our study reveals the half-metallic electronic structure with highly spin-polarized $\Delta_1$ band, which is robust against atomic disorder. Next we studied the electronic structure of  Mn$_2$RuAl/MgO and Mn$_2$RuGe/MgO heterojunctions. We found that the MnAl- or MnGe-terminated interface is energetically more favorable compared to the MnRu-terminated interface. Interfacial states appear at the Fermi level in the minority-spin gap for the Mn$_2$RuGe/MgO junction. We discuss the origin of these interfacial states in terms of local environment around each constituent atom. On the other hand, in the Mn$_2$RuAl/MgO junction, high spin polarization of bulk Mn$_2$RuAl is preserved independent of its termination.

\end{abstract}

\keywords{Density functional theory, Heusler alloys, Half-metallic, Density of states}

\maketitle

\section{Introduction} 
Half-metallic ferromagnets have attracted much attention in the last few decades because of their potential applications in the spintronics devices.\cite{Miura-PRB-2011,Miura-JPCM-2009,Miura-prb,PRL-Hulsen,JAP-Ishikawa-2008,APL-Liu-2012,PRM-Tsuchiya} In particular, Co-based Heusler alloys play an important role as electrode materials of magnetic tunnel junctions (MTJ), since many of these alloys show high spin polarization at the Fermi level ($E_\mathrm{F}$) and high Curie temperature ($T_\mathrm{C}$). Another important feature of these Co-based Heusler alloys is their robustness of spin polarization against atomic disorder.\cite{prb-miura-2004,jap-miuara-2004,jap-miura-2006,prb-shigeta-2018,JPCM-umetsu-2019}

There have been a significant amount of experimental studies and first-principles calculations on Co-based Heusler alloy/MgO heterojunctions.\cite{Gui-apl-2011,kammer-apl-2004,Sakuraba-apl-2006,lakhan-apl-2018,Saito-PRB-2010,Miura-jpcm-2007} Among them, Co$_2$MnSi/MgO heterojunction is one of the most studied systems. Although Co$_{2}$MnSi is  half-metallic in its bulk form, the half-metallicity vanishes  when it forms heterojunction with MgO, i.e. interfacial states appear at the $E_\mathrm{F}$ in the minority-spin energy gap.\cite{Saito-PRB-2010,Miura-jpcm-2007} 
Mavropoulos {\it et al.} argued that the interfacial states can act as spin-flipping center and eventually deteriorates the spin-filtering property of the heterojunction.\cite{Mavropoulos-PRB-2005} Thus, it is important to preserve the half-metallic nature at the interface of Heusler-alloy/MgO heterojunctions.

Another issue with the  Co$_2$MnSi/MgO heterojunction is the lattice mismatch of about 5$\%$.\cite{Miura-prb} Previous studies suggest that the  lattice mismatch could result misfit dislocation at the junction and degrades the spin-transport properties.\cite{Bonell-IEEE, Bonell-PRB} Recently, Kunimatsu {\it et al.} clarified the role of dislocation density in the reduction of the tunnel magneto resistance (TMR) ratio for the Co$_3$Mn/MgO MTJ.\cite{Kunimatsu-apex} All these previous studies suggest, for a better spin-transport property, lattice matching between MgO and Heusler alloy is also an important prerequisite. 
Recently, it  was reported that the replacement of one Co of Co$_2$-based Heusler alloys by an isoelectronic Ir could reduce the interfacial lattice mismatch to less than 1$\%$.\cite{troy-jmmm-2020}  For the Heusler-alloy/MgO (001) heterojunction, there is perfect lattice matching since the Heusler alloy has the lattice parameter of 5.95 \AA.

Besides the Co$_2$-based Heusler alloys, Mn$_2$CoZ (Z=s,p elements) with Hg$_2$CuTi-type inverse Heusler structure systems have been predicted to have high spin polarization.\cite{Meinert-jpcm}  Because of the inverse Heusler structure, the Heisenberg exchange coupling constants between two Mn atoms become stronger, which results in high $T_\mathrm{C}$, as reported by Meinert {\it et al.}\cite{Meinert-jpcm}  However, as both Mn and Co are $3d$ elements, the lattice parameter is on the shorter side of our requirement on the  compatibility with MgO.

Here, we tried to seek for the Heusler alloys, which have high spin polarization in bulk form and its lattice parameter is around 5.95 \AA, such that the interfacial lattice mismatch with MgO becomes as small as possible. Possible candidates are Mn$_2$RuZ (Z= Al, Si, Ga, Ge). The lattice parameters of these alloys are  well matched with MgO for construction of heterojunction along (001) direction. Previous first-principles studies suggest nearly half-metallic electronic structure in their bulk form.\cite{Ghosh-phys-scr, Shimosakaida-mater-trans, Song-jmmm-2017,Gupta-jalcom-2013,Ling-jmmm-2015}
These motivate us to investigate the Mn$_2$RuZ/MgO (001) heterostructures.   Endo {\it et al.}\cite{Endo-jalcom-2012} first reported Mn$_2$RuSi to have spin-glass like magnetic structure. However, later on there are several reports on the same material, suggesting ferrimagnetic ground state.\cite{ Shimosakaida-mater-trans, Song-jmmm-2017,Gupta-jalcom-2013} Mn$_2$RuGa and Mn$_2$RuGe have also been experimentally prepared in their respective bulk phases.\cite{Ling-jmmm-2015} Apart from Mn$_2$Ru$_x$Ga/MgO (001)\cite{Borisov-apl-2016}, there have been no existing reports on Mn$_2$RuZ/MgO heterostructures.  

In the spin-dependent transport properties of MTJ with Heusler-alloy electrodes and single crystalline MgO barrier,
the presence of spin-polarized $\Delta_1$-band for the Heusler alloy plays an important role.\cite{butler} Therefore, before studying the  Heusler-alloy/MgO (001) heterojuntion, it is of extreme interest to discuss the symmetry of the bands crossing the $E_\mathrm{F}$  along the $\Delta$ axis.

Here, we focus on the electronic and magnetic properties of  Mn$_2$RuZ/MgO (001) heterojunctions.  Firstly, we carried out first-principles calculations of the electronic structure for these bulk systems, and then we investigate the Mn$_2$RuZ/MgO (001) heterojunctions.

\section{Method}  
Geometry optimization of bulk structures has been carried out using Vienna Ab Initio 
Simulation Package 
(VASP)\cite{VASP} in combination
with the projector augmented wave (PAW) 
method.\cite{PAW}
For the exchange correlation functional, we use generalized gradient approximation (GGA) as parameterized by Perdew, Buke and Ernzerhof.\cite{PBE} A cutoff energy of 
500\,eV has been used for the planewaves. The final energies have 
been calculated with a $k$-mesh of 
16$\times$16$\times$16 for the bulk phases. 
The tolerances in total energy and the force of our calculations are 10 $\mu$eV and 10 meV/\AA, respectively.

We constructed the heterojunction with eleven layers of Mn$_2$RuZ on top of five layers of MgO along (001) direction. The out of plane lattice parameter was relaxed, whereas the in plane lattice parameter was fixed to that of the optimized lattice parameter of bulk Mn$_2$RuZ system. It is to be mentioned that the in plane lattice parameter is
 $a/\sqrt{2}$, where $a$ is the optimized lattice parameter for bulk Heusler system. For the geometry optimization of the junction along $c$-axis we use a $k$-mesh of 10$\times$10$\times$1, while the electronic structure and magnetic properties of these heterojunctions have been obtained using a $k$-mesh of 16$\times$16$\times$2.

Green's function based 
spin-polarized relativistic 
Korringa-Kohn-Rostoker method (SPR-KKR) has been used as implemented in the 
SPR-KKR programme package\cite{sprkkr} in order 
to calculate Heisenberg exchange coupling constant within a real space approach proposed by Liechtenstein {\it et al.}\cite{licechenstein} Thereafter, we calculate $T_\mathrm {C}$ within a mean-field approximation (MFA) for the bulk systems. A full potential method has been used for the self-consistent-field calculations. 
For the determination of $E_\mathrm{F}$, we used Lloyd's formula.\cite{lloyd1, lloyd2}  
In this method also we use the GGA exchange correlation functional.\cite{PBE} For all atoms, the angular momentum cut-off  $l_\mathrm{max}$ was restricted to two and the Brillouin zone was sampled with a $k$-mesh of 26$\times$26$\times$26. 
The calculations for the disordered phases have been carried out using a coherent potential approximation. 

\section{Results and Discussion}
\begin{table*}[hbt!]

\centering
Table~1. Calculated bulk properties of Mn$_2$RuZ.\footnote{Comparison with experiments or previous calculations, wherever data are available\\ 
$^{b}$Ref.\onlinecite{Ghosh-phys-scr} (theoretical result)
$^{c}$Ref.\onlinecite{Ling-jmmm-2015} (theoretical and experimental result)
$^{d}$Ref.\onlinecite{Hori-apa-2002} (experimental result)
$^{e}$Ref.\onlinecite{Betto-aip-adv-2016} (experimental result)
$^{f}$Ref.\onlinecite{Galanakis-jap-2014} (theoretical result)
$^{g}$Ref.\onlinecite{Endo-jalcom-2012} (experimental result)
$^{h}$Ref.\onlinecite{Shimosakaida-mater-trans} (theoretical result)
} 
\begin{tabular}{|c|c|c|c|c|c|c|c|c|}
\hline Material & $a$  & $\mu_{total}$ & $\mu_{\mathrm Mn_A}$ & $\mu_{\mathrm Mn_B}$ & $\mu_\mathrm {Ru}$ & $\mu_{Z}$& $T_\mathrm C$ & $P$ (\%)\\
&(\AA)&($\mu_{B}$)&($\mu_{B}$)&($\mu_{B}$)&($\mu_{B}$)&($\mu_{B}$)&(K)& \\
\hline Mn$_2$RuAl & 5.94 &1.00 & -2.15& 3.04 & 0.06 & -0.02 &670&99\\
&5.95$^b$&1.01$^b$&-2.18$^b$&3.07$^b$&0.06$^b$&0.02$^b$&&96$^b$\\
\hline Mn$_2$RuGa & 5.96 &1.02 & -2.27& 3.12 & 0.08 & 0.00 &693 &96\\
&6.00$^c$&1.03$^c$,1.15$^d$&-2.24$^c$&3.04$^c$&0.20$^c$&0.02$^c$& 460$^d$,550$^e$, 670$^f$&\\
\hline Mn$_2$RuSi & 5.79 &2.00 & -0.85& 2.68 & 0.06 & -0.07 &313& 100\\
&5.826$^g$, 5.801$^h$ &2.002$^h$&-0.809&2.658&0.068&0.027&&\\
\hline Mn$_2$RuGe & 5.90 &2.00 & -1.16& 2.92 & 0.12 & -0.01 &489&97\\
&5.92$^c$, 5.91$^c$&1.55$^c$, 2.00$^c$&-1.12$^c$&2.84$^c$&0.24$^c$&0.06$^c$&303$^c$&\\

\hline
\end{tabular}   
\end{table*}
\begin{figure}[h]

\includegraphics[width=0.8\textwidth]{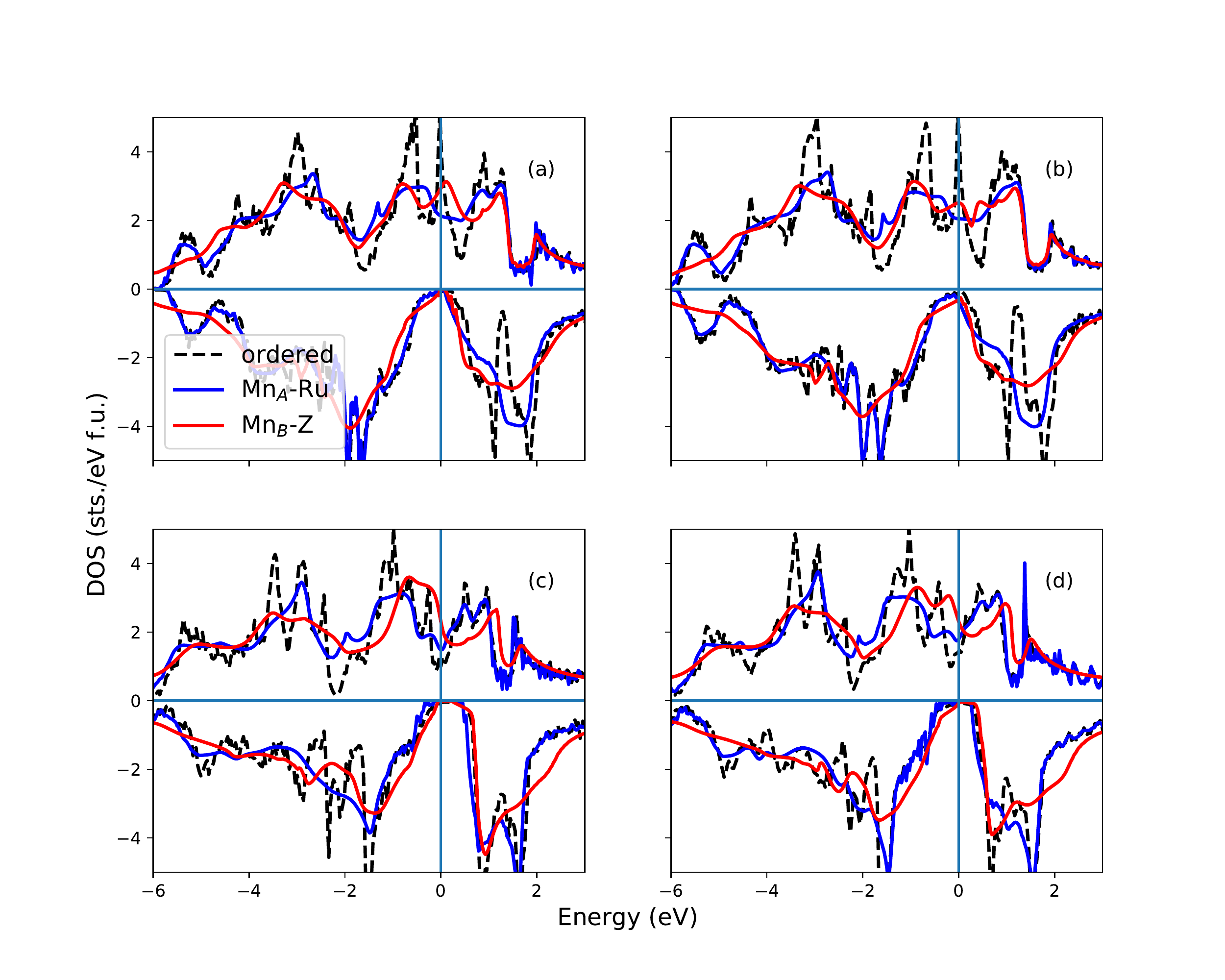}
\caption
{(Color online) The spin resolved and total density of states (DOS) for the ordered, full random swapping between Mn$_A$ and Ru, and between Mn$_B$ and $Z$, ($Z$=Al, Ga, Si, Ge depending on the systems) for (a) Mn$_2$RuAl, (b) Mn$_2$RuGa, (c) Mn$_2$RuSi and (d) Mn$_2$RuGe, respectively. The Fermi level is at 0 eV.} 

\end{figure}

\begin{figure}[h]

\includegraphics[width=0.8\textwidth]{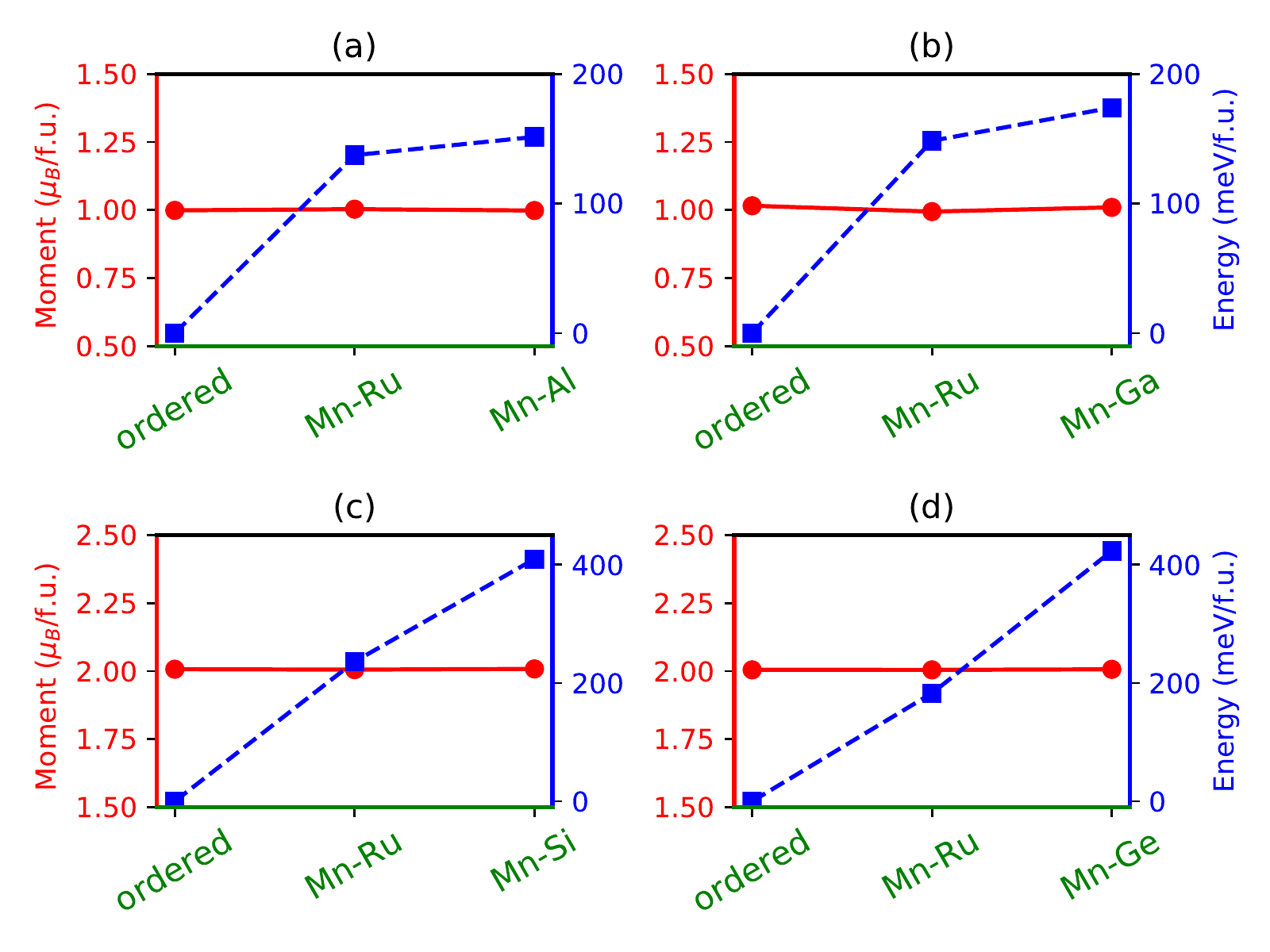}
\caption
{(Color online) Comparison of total magnetic moment and total energy of the ordered phase, Mn(Mn$_A$)-Ru full swapping, Mn(Mn$_B$)-Z full swapping for (a) Mn$_2$RuAl, (b) Mn$_2$RuGa, (c) Mn$_2$RuSi, (d) Mn$_2$RuGe.} 

\end{figure}

\subsubsection{Bulk phase}

The systems studied here possess XA-type inverse Heusler structure. In the bulk structure there are four interpenetrating face-centered cubic sublattices with their centers at (0, 0, 0), (0.25, 0.25, 0.25), (0.5, 0.5, 0.5), and (0.75, 0.75, 0.75). We name these sublattices as $A$, $B$, $C$, and $D$, respectively. For Mn$_2$RuZ (Z= Al, Ga, Si, and Ge), $A$, $B$, $C$, and $D$ sublattices are occupied by Mn (Mn$_A$), Mn (Mn$_B$), Ru, and Z atoms. In the  ground state magnetic configuration Mn$_A$ has an anti-parallel spin alignment with respect to all other magnetic atoms of the systems, resulting a ferrimagnetic ground state. In Table 1 we have summarized the bulk properties, i.e., lattice parameter ($a$), total magnetic moment ($\mu_{tot}$), atom resolved magnetic moment, spin polarization ($P$) at the  $E_\mathrm{F}$, and the $T_\mathrm {C}$. These data also have been compared with the existing literature and found to be in well agreement with these reports.

\begin{figure}[h]
\includegraphics[width=0.8\textwidth]{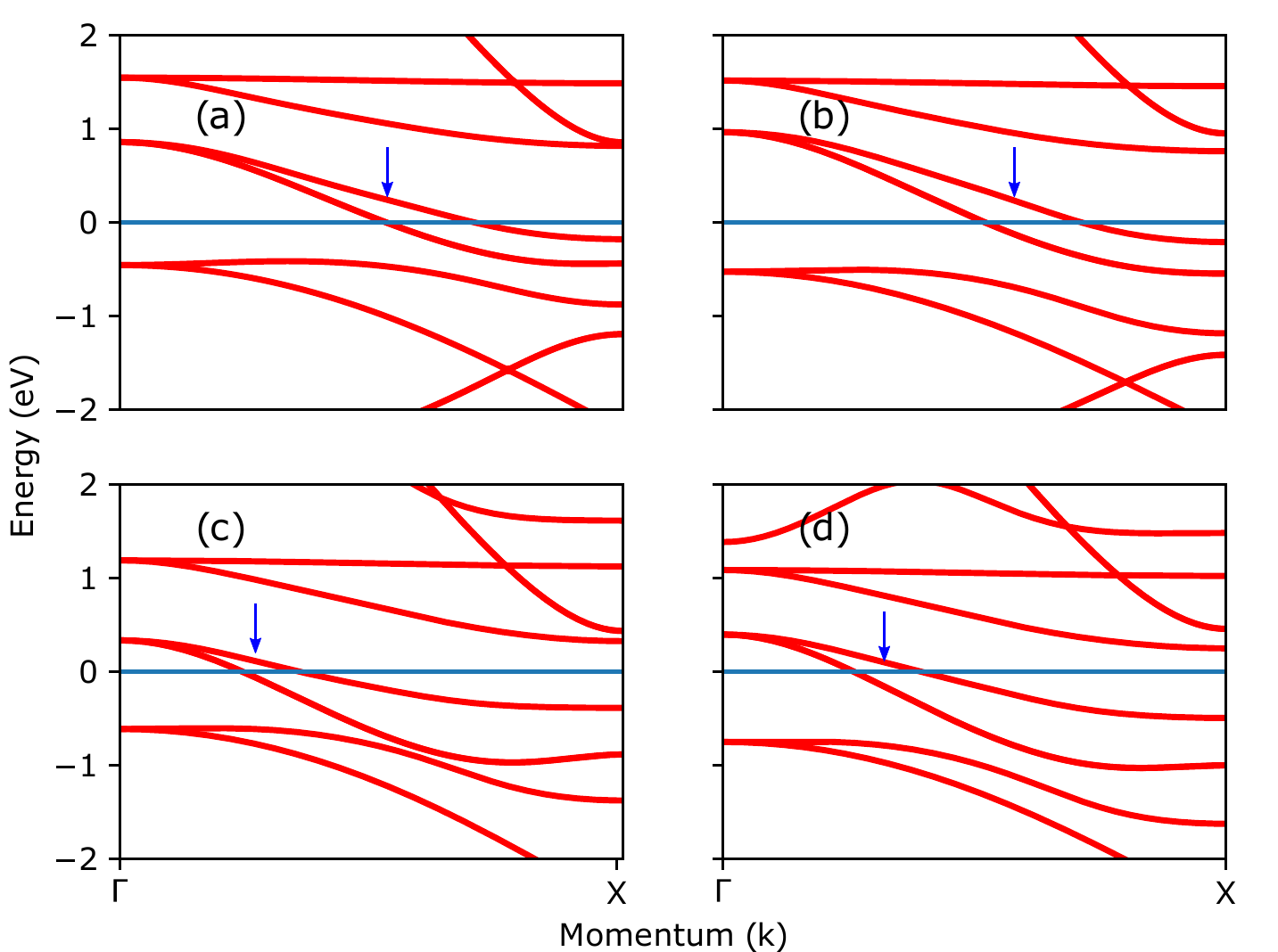}
\caption
{ Majority spin band structure along $\Delta$ direction for: (a) Mn$_2$RuAl, (b) Mn$_2$RuGa, (c) Mn$_2$RuSi and (d) Mn$_2$RuGe, respectively. Arrows indicate the band with $\Delta_1$ symmetry. The Fermi level is at 0 eV.} 

\end{figure}

For all these alloys except Mn$_2$RuSi, the lattice parameter is in the range from 5.90 \AA{} to 5.95 \AA, which could be beneficiary in the formation of Mn$_2$RuZ/MgO (001) heterojuction with reasonable interfacial lattice matching as mentioned earlier. The total magnetic moment for all these systems have integer value or closed to integer value. For Z=Al or Ga the total magnetic moment is equal to unity and it becomes 2.00 $\mu_{B}$ for  Z =Si or Ge. This is consistent with the Slater-Pauling rule.\cite{Galanakis-prb} From the viewpoint of applications to spin-transfer torque (STT) magnetoresistive random access memories (MRAM), ferrimagnets possessing lower magnetization, such as $D$0$_{22}$-type Mn$_3$Ga and related materials\cite{balke-apl,mizukami-prb,mizukami-prl}, are beneficial to reduce the magnetization switching current.

We found that $T_\mathrm{C}$  is higher for Mn$_2$RuAl and Mn$_2$RuGa compared to Mn$_2$RuSi and Mn$_2$RuGe. The higher $T_\mathrm{C}$ for the first two systems could be explained by the larger absolute values of magnetic moments of Mn$_A$ and Mn$_B$, which causes stronger exchange coupling between Mn$_A$ and Mn$_B$. It is observed that the calculated $T_\mathrm{C}$ is higher than the experimental values. The  overestimation of $T_\mathrm{C}$ within the MFA has been reported by several groups.\cite{prb-sasioglu-2005,prb-sasioglu-2005-2,prb-rusz} The MFA does not consider any spin fluctuations, and thus it neglects the energy gain owing to the short-ranged magnetic order in the paramagnetic phase. This is the reason why the MFA usually overestimates $T_\mathrm{C}$.

The spin polarization of these systems are equal to or very close to 100\%, signifying the nearly half-metallic nature of these systems. In Figure 1, we show the spin-polarized density of states (DOS) of each system in its ordered phase (XA) and its possible disordered phases which results from full random swapping between Mn$_B$-Z and Mn$_A$-Ru atoms, and results in a $L2_1$ structure. From Figure 1, it is obvious that the spin polarization at the $E_\mathrm{F}$ is of robust nature against the considered disordered phases. The robustness of spin-polarization against atomic disorder has been reported for Co-based Heusler alloys.\cite{prb-miura-2004,jap-miuara-2004,jap-miura-2006,prb-shigeta-2018,JPCM-umetsu-2019}

Figure 2 shows the comparison of stability of the disordered systems ($L2_1$) with respect to that of their ordered configuration ($XA$). In all the cases, ordered configuration is the most stable configuration and the Mn-Z disorder is the least stable configuration. Note that energy of Mn-Z disordered phase with respect to the ordered phase is almost twice for Z=Si or Ge (409 meV/f.u., 423 meV/f.u.) compared to the Z=Al or Ga (151 meV/f.u., 174 meV/f.u.). We know that Al and Ga are metal in bulk form, their electronegativity is closer to that of Mn.  
 On the other hand, bulk Si and Ge are semiconductor, their electronegativity is significantly different from Mn. It could results in antisite disorder of Mn and Si/Ge to be energetically more unfavorable in Mn$_2$RuSi/Mn$_2$RuGe than that of Mn and Al/Ga disorder in  Mn$_2$RuAl/Mn$_2$RuGa.  
The total magnetic moment of the system shows robust nature against the considered disordered cases, which arises from the robustness of the spin polarization against atomic disorder.

As it has been already mentioned, it is very important to discuss the bandstructure along (001) direction ($\Delta$ line along the Brillouin zone) and consider the symmetry of bands crossing the $E_\mathrm{F}$. As the systems are half-metallic, we only include the discussion for the majority-spin channel around the  $E_\mathrm{F}$. As shown in Fig. 3, we find that for all the Mn$_2$RuZ (Z=Al, Ga, Si, Ge) systems, the majority spin band with $\Delta_1$ symmetry (marked by arrow) crosses the  $E_\mathrm{F}$. A clear shift of the $\Delta_1$ band towards the higher binding energy side could be observed in Mn$_2$RuSi and  Mn$_2$RuGe, compared to the rest of the two systems, namely,  Mn$_2$RuAl and  Mn$_2$RuGa. This shift  is a direct consequence of the difference in valence-electron numbers which is larger by one for the cases of Z=Si or Ge.

From the above discussion on the bulk phases, we find that these systems could be interesting in further study regarding its junction with MgO along (001) direction.

\subsubsection{Heterojunction with MgO}

In this section we consider the heterojunctions of Mn$_2$RuZ (Z=Al and Ge) with MgO. The interfacial lattice mismatch is about -0.2\% and -0.9\% for the  Mn$_2$RuAl/MgO (001) and  Mn$_2$RuGe/MgO (001) heterojunctions, respectively. Here, we consider the configuration, in which the O-atoms of MgO are located on the top of (Mn (Mn$_A$), Ru) or (Mn (Mn$_B$), $Z$) atoms depending on the MnRu- or MnZ-terminated interface. 

\begin{figure}[h]
\includegraphics[width=1\textwidth]{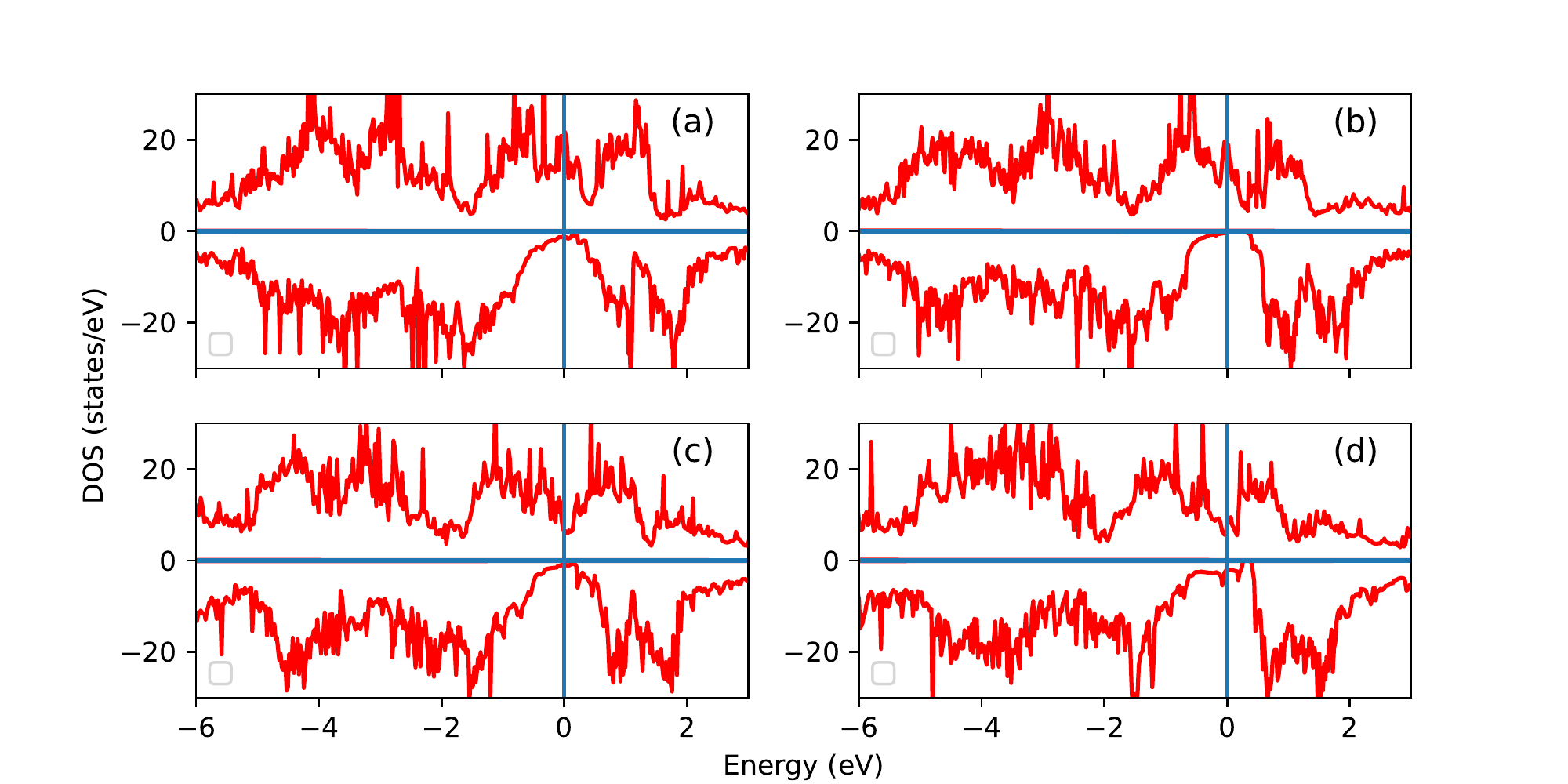}
\caption
{ The spin resolved and total density of states (DOS) : (a) Mn$_A$Ru-terminated, (b)  Mn$_B$Al-terminated interface for Mn$_2$RuAl/MgO (001); (c) Mn$_A$Ru-terminated, (d)  Mn$_B$Ge-terminated interface for Mn$_2$RuGe/MgO (001). The Fermi level is at 0 eV.} 

\end{figure}

\begin{figure}[h]
\includegraphics[width=1\textwidth]{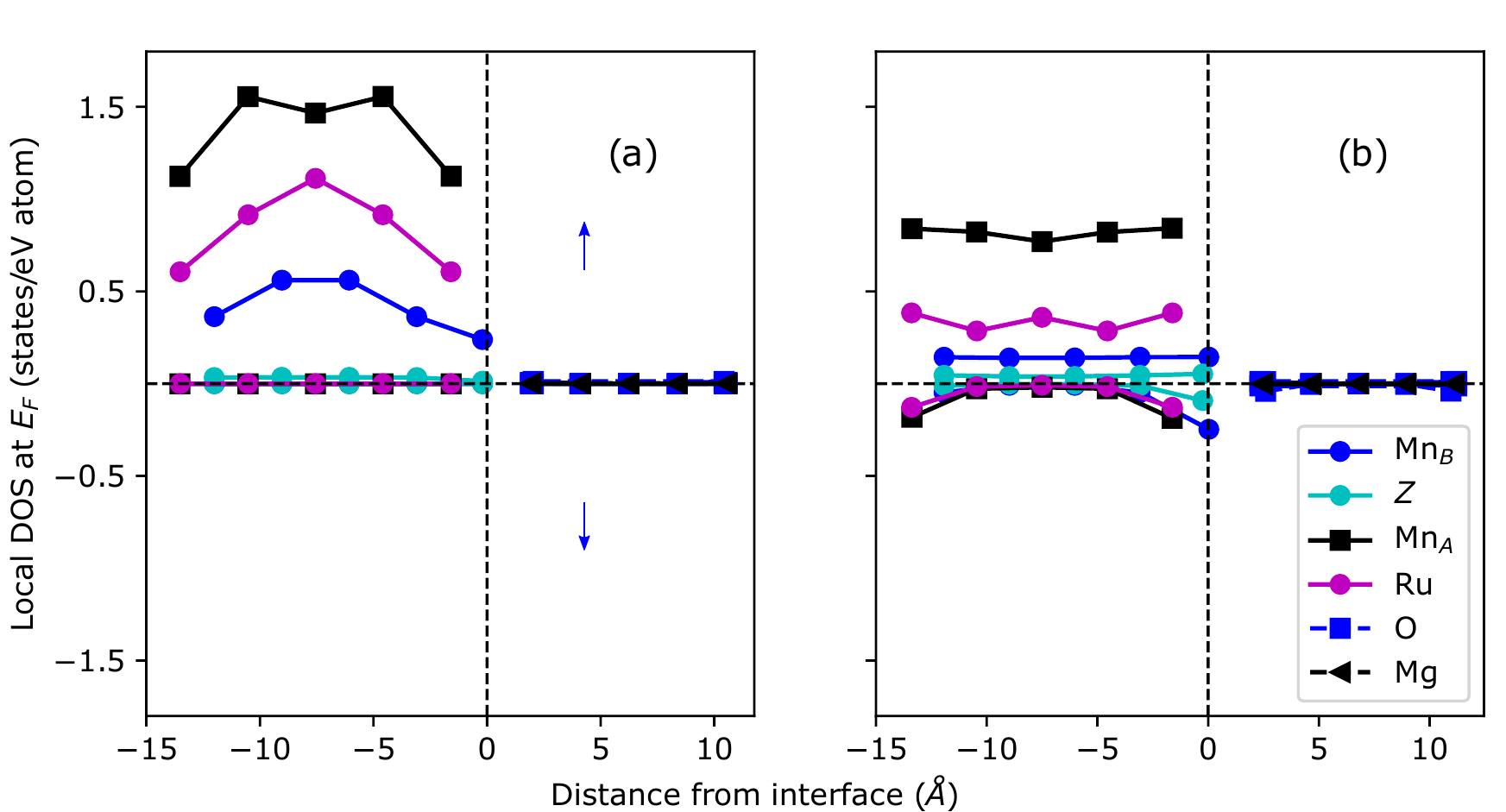}
\caption
{(Color online)  Local DOS of each atom at $E_\mathrm{F}$ for each atom plotted as a function of distance for: (a) Mn$_2$RuAl/MgO (001), (b) Mn$_2$RuGe/MgO (001), respectively.} 

\end{figure}

\begin{figure}[h]
\includegraphics[width=1\textwidth]{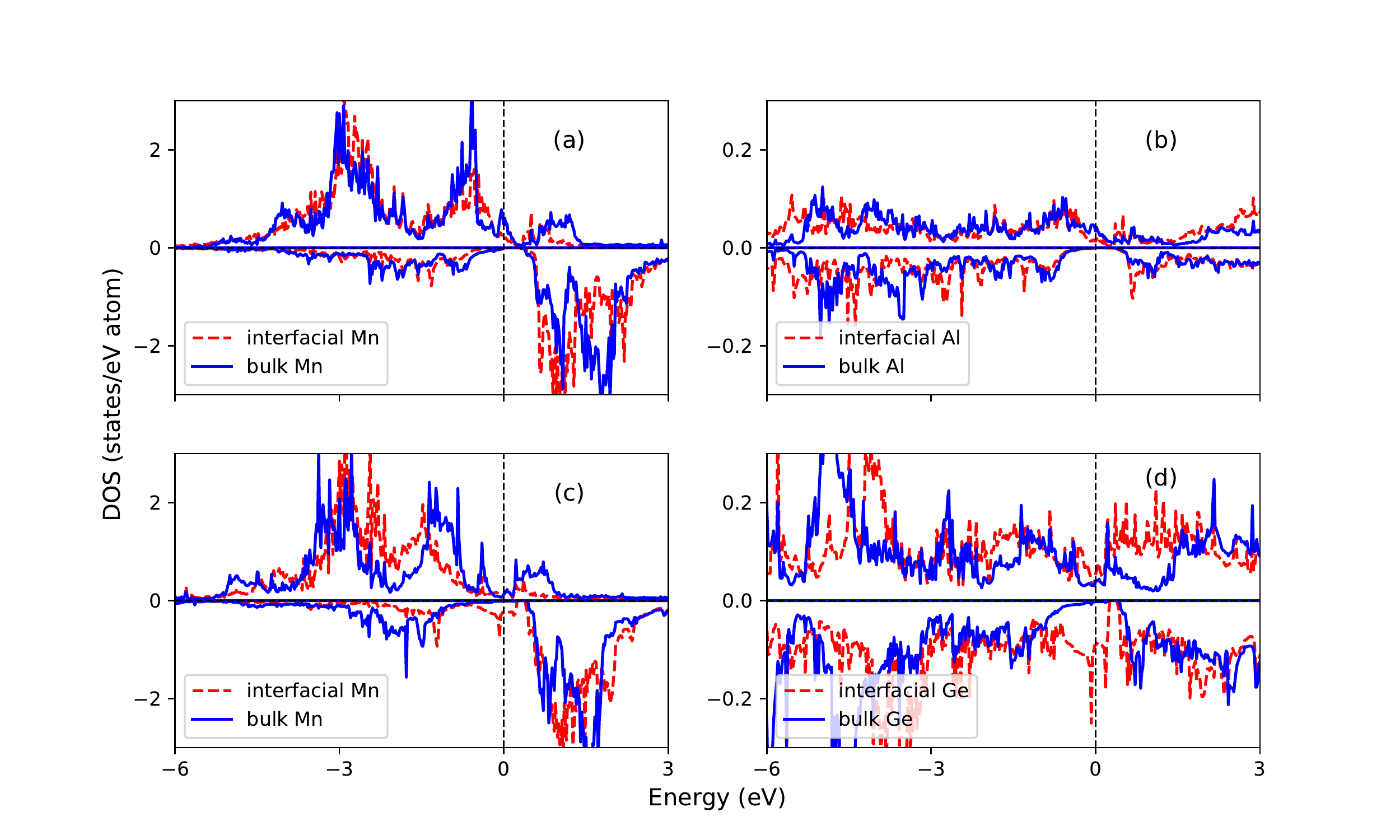}
\caption
{(Color online)  Dashed (red) lines show the local DOS of interfacial (a) Mn, (b) Al of Mn$_2$RuAl/MgO (001) heterojunction; (c) Mn, (d) Ge of Mn$_2$RuGe/MgO (001) heterojunction. In each case the local DOSs of the bulk region is also shown by the solid  (blue) lines, for comparison. The Fermi level is at 0 eV.} 

\end{figure}
Figure 4 shows the spin-polarized DOS for the Mn$_2$RuAl/MgO (001) and Mn$_2$RuGe/MgO (001) heterojunctions, for both Mn$_A$Ru- and Mn$_B$Z-terminated interfaces. Note that for all the cases the high spin polarization at the $E_\mathrm{F}$ is preserved for both of the Mn$_2$RuAl and Mn$_2$RuGe systems in their heterojunctions with MgO. The spin polarization is 100\% for the Mn$_B$Al-terminated interface of the Mn$_2$RuAl/MgO heterojunction.  This could be effective in the suppression of tunneling conductance in the antiparallel magnetization of Mn$_2$RuAl/MgO-based MTJ.\cite{Miura-prb}
Although, for the other cases, there is a  decrease of spin-polarization at the interface, still spin polarization is considerably high, which are 89\%, 76\% and 57\% for the Mn$_A$Ru-terminated interfaces of  Mn$_2$RuAl/MgO (001) and Mn$_2$RuGe/MgO (001) and Mn$_B$Ge-terminated interface of Mn$_2$RuGe/MgO (001), respectively.
It is worth investigating the origin of the appeared states at the  $E_\mathrm{F}$ in the minority-spin gap at the interface.

We compared the stability of interfacial structure in terms of the formation energy and found that Mn$_B$Z-terminated interface is energetically more favorable than the Mn$_A$Ru-terminated interface. This observation is quite consistent with the previous reports on Co$_2$MnSi/MgO, Co$_2$CrAl/MgO and related systems.\cite{Miura-prb,troy-jmmm-2020}
Hereafter, we focus on the more detailed discussion for Mn$_B$Z-terminated interface.

Figures 5(a) and 5(b) show the local DOS for each atom at the  $E_\mathrm{F}$ as a function of distance from the junction. It is to be noted that there are no states at the  $E_\mathrm{F}$ arising from the minority-spin channel in Mn$_2$RuAl/MgO heterojunction, leading to a half-metallic interface. However, for the  Mn$_2$RuGe/MgO heterojunction states appear in the minority-spin channel from the interfacial Mn$_B$ and Ge atoms.

The interfacial region plays a crucial role in determining the electronic and magnetic properties of the heterojunctions. So, it is important to discuss the electronic structure of the interfacial atoms.
Figure 6(a) and 6(b) show the DOS of the interfacial Mn and Al atoms for Mn$_2$RuAl/MgO, and 6(c) and 6(d) show the same of the interfacial Mn and Ge atoms for Mn$_2$RuGe/MgO. In each case, we compare the DOS of the interfacial atom with the bulk region of the junction to have a better understanding of modifications of electronic structure owing to the change in local environment around each atomic site. We found that for both the systems the DOS of the interfacial Mn atom has been shifted toward the higher binding energy side compared to that of the bulk Mn. This leads to increase of the interfacial Mn moment (3.52 $\mu_{B}$) compared to that of the bulk Mn atom (3.02 $\mu_{B}$) in  Mn$_2$RuAl/MgO heterojunction. In Mn$_2$RuGe/MgO the values of interfacial and bulk Mn moments are 3.72 $\mu_{B}$ and 2.92 $\mu_{B}$, respectively. This enhancement of the magnetic moment at the interface is quite consistent with the previous report on other heterojunctions of Co-based Heusler alloys and MgO.\cite{Miura-prb,troy-jmmm-2020} It has been argued that the charge transfer from the minority-spin to the majority-spin band causes the enhancement of magnetic moment of the interfacial Mn atoms.

\begin{figure}[h]
\includegraphics[width=0.5\textwidth]{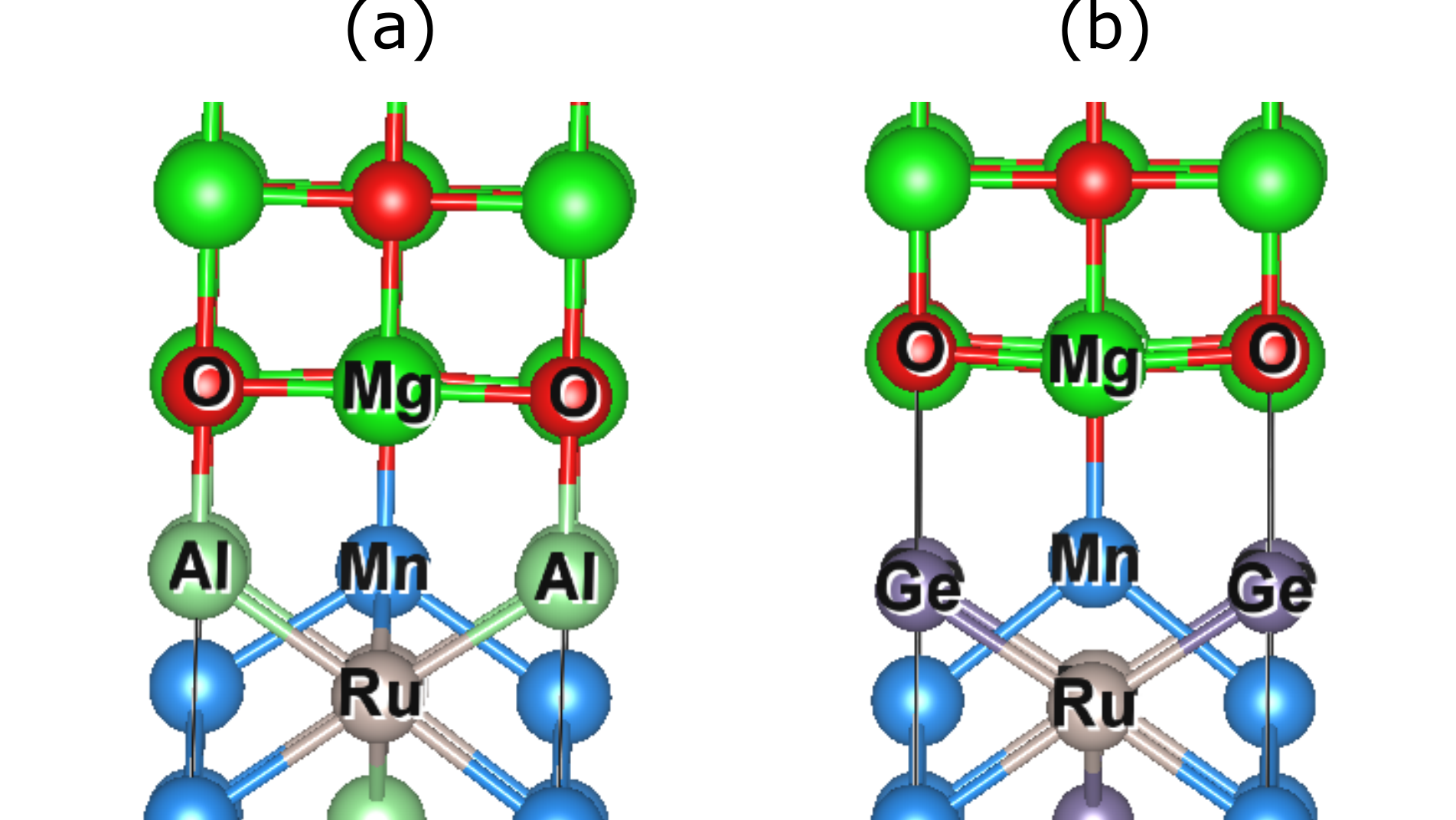}
\caption
{ The interfacial structures of (a) Mn$_2$RuAl/MgO(001) and (b) Mn$_2$RuGe/MgO(001) junctions.} 

\end{figure}

Now we discuss the underlying reason behind the appearance of the interfacial states in the Mn$_2$RuGe/MgO heterojunction whereas  Mn$_2$RuAl/MgO heterojunction is half-metallic.  It is important to consider the strength of interatomic bonding at the interface. Figure 7(a) and 7(b) represent the interfacial structure of  Mn$_2$RuAl/MgO and Mn$_2$RuGe/MgO heterojunctios, respectively. We found that the bond length between Mn$_B$ and O (2.23 \AA) is qute similar to that of the Al and O (2.10 \AA) in Mn$_2$RuAl/MgO junction. However, the interfacial structure is quite different in Mn$_2$RuGe/MgO junction, in which the the Mn$_B$-O and Ge-O bond lengths are  2.30 \AA, and 2.84 \AA, respectively. This weak bonding between Ge and O could cause appearance of electronic states at the  $E_\mathrm{F}$, and eventually diminishes the half-metallicity at the interface. This case is quite similar with the previous report on CoIrMnSi/MgO heterjunction.\cite{troy-jmmm-2020}

 \section{Conclusion}
Based on first-principles calculation, we study Mn$_2$RuZ (Z=Al, Ga, Si, Ge) in their bulk forms. All these systems have high spin polarization with  $\Delta_1$ band across the $E_\mathrm{F}$. The spin polarization of these systems is robust in nature against the atomic disorder. Among all these systems, Mn$_2$RuAl possess relatively high $T_\mathrm{C}$ of 670 K.
We found that high spin polarization at the $E_\mathrm{F}$ is maintained in Mn$_2$RuAl/MgO (001) heterojunctions for both MnRu- and MnAl-terminated interfaces, i.e. 89\% and 100\%, respectively. In particular, the MnAl-terminated interface is energetically more favorable. In conclusion, Mn$_2$RuAl is a promising candidate for electrode materials in MgO-based MTJ since it possesses desirable properties mentioned above. 
\section{Acknowledgements}  
 The authors are grateful to S. Mizukami, A. Hirohata, T. Tsuchiya, and T. Ichinose for valuable discussion. TR thanks T. Kanomata for his valuable suggestions. This work was partially supported by JST CREST (No. JPMJCR17J5) and by CSRN, Tohoku University.

{}

\clearpage

\end{document}